\documentclass[twocolumn,aps,prd,reprint,showpacs,amsmath,amssymb]{revtex4-1}
\usepackage{epsfig}
\usepackage{graphicx}
\usepackage{dcolumn}
\usepackage{bm}
\usepackage{bookmark}
\usepackage{color}
\usepackage{ulem}

\begin{document}
\title{Kibble-Zurek mechanism in quantum link model}
\author{Yao-Tai Kang$^1$}
\author{Chung-Yu Lo$^1$}
\author{Shuai Yin$^2$} \email{zsuyinshuai@163.com}
\author{Pochung Chen$^1$} \email{pcchen@phys.nthu.edu.tw}
\affiliation{$^1$Department of Physics, National Tsing Hua University, Hsinchu 30013, Taiwan}
\affiliation{$^2$Institute for Advanced Study, Tsinghua University, Beijing 100084, China}

\date{\today}

\begin{abstract}
We study the driven critical dynamics of the quantum link model, whose Hamiltonian describes the one-dimensional $U(1)$ lattice gauge theory.
We find that combined topological defects emerge after the quench and they consist of both gauge field and matter field excitations.
Furthermore, the ratio of gauge field and matter field excitation is $1/2$ due to the constraint of the Gauss' law.
We show that the scaling of these combined topological defects satisfies the usual Kibble-Zurek mechanism.
We verify that both the electric flux and the entanglement entropy satisfy the finite-time scaling theory in the whole driven process.
Possible experimental realizations are discussed.
\end{abstract}

\maketitle

\section{\label{intro}Introduction}
Lattice gauge theories (LGTs) play important roles in fathoming a variety of strongly correlated systems~\cite{Wen,Fradkin,Sachdevbook,Degrand,Rothe}. In particle physics, where gauge theories appear as fundamental degrees of freedom, the LGT provides a non-perturbative approach to continuum gauge field theories like quantum electrodynamics (QED) or quantum chromodynamics (QCD)~\cite{Degrand,Rothe}. In condensed matter physics, the LGT usually arises as an emergent low-energy theory, which is constructed by rewriting the Hamiltonian in terms of new collective degrees of freedom~\cite{Wen,Fradkin}. Prominent examples include the LGT descriptions of the high-$T_c$ superconductors~\cite{Baskaran,Affleck,Lee}, the quantum frustrated magnets~\cite{Zhou}, and topological phases~\cite{WenR}. Moreover, recent rapid advances have made it possible to perform quantum simulation of Abelian and non-Abelian LGTs with ultracold atoms in optical lattices, and they can be engineered to imitate certain phenomena in high-energy physics~\cite{Zohar2012,Banerjee2013,Zohar2013}.

On the other hand, developing effective theories to describe the non-equilibrium phenomena in quantum systems is of central significance in condensed matter physics and ultracold atom physics~\cite{revqkz1,revqkz2}. Among them, the driven critical dynamics under an external driving stands out remarkably. This is in part stimulated by its potential application in quantum simulation and quantum computer~\cite{revqkz1,revqkz2}. Theoretically, the Kibble-Zurek mechanism (KZM) provides a general description of the generation of the topological defects and the scaling of their number after the quench~\cite{Kibble1,Zurek1}. While the KZM was originally proposed in classical phase transitions, it has been generalized into the quantum cases~\cite{qkz1,qkz2,qkz3,qkz4,qkz5,qkz6}. Moreover, the quantum KZM has been verified in various experiments~\cite{Ulm,Pyka}. Besides the scaling of the topological defects, recent theoretical and experimental studies also pay close attentions to the full scaling in the whole driven process~\cite{qkz7,qkz8,qkz9,Chandran,qkz10,Yin,Gerster}. For example, the finite-time scaling (FTS) has been verified in the driven critical dynamics of the Rydberg atomic systems~\cite{Keesling}. Furthermore, the FTS has been employed to numerically detect the critical properties in both classical~\cite{Zhong1,Zhong2} and quantum phase transitions~\cite{qkz10,Hu,HuangRZ}.

Inspired by the great progress in experiments, the real-time dynamics of LGTs has attracted enormous attentions recently. These include the glassy dynamics of the many-body localization state induced by superselection of gauge sectors~\cite{Smith2017,Smith2018,Brenes}, the fermion production and string breaking in cold atom simulators~\cite{Martinez,Banerjee,Hebenstreit,Spitz}, and the dynamical phase transitions after a sudden quench in the Schwinger model~\cite{HuangYP,Zache}. However, it was shown that the number of the topological defects may not satisfy the usual KZM in some local gauge invariant systems due to the gauge field fluctuation~\cite{Hindmarsh,Stephens}. This motivates us to further explore the driven critical dynamics in LGTs.

In this work, we explore the driven critical dynamics of the one-dimensional ($1$D) quantum link model (QLM)~\cite{Chandrasekharan,Brower}. This model approximates the $1$D Schwinger model, which describes the $1$D $U(1)$ gauge theory under certain static electric field~\cite{Schwinger,Coleman}, by replacing the infinite continuous Hilbert space of the gauge field with a discrete one. For this model, a phase transition associated with parity ($P$) and charge conjugate ($C$) symmetry breaking can happen by tuning the chemical potential of the matter field~\cite{Rico}. In passing, we note that due to the constraint of the Gauss' law, the topological defects in this model consist of both the gauge field excitation and the matter field excitation. Moreover, as it will be shown later, the ratio of their numbers is $1/2$. First, we detect the topological defects of this model after quench by linearly changing the chemical potential of the mass term from symmetric phase into symmetry breaking phase. We find that the scaling of topological defects satisfies the usual KZM. We then show that the topological defects are combination of the excitations in both gauge field and matter field by calculating their ratio. Next, we study the full scaling behavior in the whole driven process. We show that both the flux and the entanglement entropy obey the FTS theory. Since the real-time dynamics in LGTs has been realized in experiments, it is expected that our present results can be examined in these kinds of systems.

The rest of the paper is organized as follows. In Sec.~\ref{schmodel}, we briefly introduce the QLM and its phase transition properties. In Sec.~\ref{combkzm}, the properties of the topological defects in the QLM are explored and the KZM is verified numerically. Furthermore, in Sec.~\ref{ftss}, we study the dynamic scaling of the flux and entanglement entropy in the whole driven process to confirm the FTS theory. Then, the universality of these scaling properties is examined in Sec.~\ref{universality}. Finally, a summary and a discussion are given in Sec.~\ref{sum}.

\section{\label{schmodel}From Schwinger model to quantum link model}
The Schwinger model describes the $1$D QED theory by coupling the fermions to the $U(1)$ gauge field.
In this work we consider the massive Schwinger model in a staggered external electric field.
Its Hamiltonian reads~\cite{Rico}
\begin{eqnarray}
\begin{aligned}
  H=&\frac{\gamma^2}{2}\sum_n[E_{n,n+1}-(-1)^n E_0]^2+\mu\sum_n(-1)^n\psi_n^\dagger\psi_n \\
      &-\kappa\sum_n\psi_n^\dagger U_{n,n+1}\psi_{n+1}+{\rm H.c.}. \label{modelham}
\end{aligned}
\end{eqnarray}
Here $\psi_n$ is the fermion annihilation operator on site $n$. $U_{n,n+1}$ is the $U(1)$ parallel transporters defined on bond $(n,n+1)$, whose corresponding electric field $E_{n,n+1}$ is overlaid by an external static field $E_0$. The staggered chemical potential $\mu$ provides a mass to the fermion~\cite{Koguts}. Furthermore, $\gamma$ measures the strength of the gauge field energy and $\gamma^2/2$ will be set to one. Finally, $\kappa$ is the coupling between the fermion field and the gauge field. The commutation relation between $E_{n,n+1}$ and $U_{m,m+1}$ is $[E_{n,n+1},U_{m,m+1}]=\delta_{n,m}U_{m,m+1}$. The gauge invariance is manifest after a local gauge transformation. Its generator is
\begin{equation}
G_n\equiv E_{n,n+1}-E_{n-1,n}-\psi_n^\dagger \psi_{n+1}+\frac{1}{2}[1-(-1)^n], \label{gene}
\end{equation}
which satisfies $[H,G_n]=0$. Consequently, states in the physical Hilbert space are constrained by the Gauss' law, which reads
\begin{equation}
G_n|\Psi\rangle=0. \label{gauss}
\end{equation}

In experiments, however, it is difficult to simulate the gauge field due to its continuum degree of freedom. An approach to tackle this problem is to convert the Schwinger model to a QLM~\cite{Chandrasekharan,Brower}, in which the gauge field is rewritten as an operator with discrete eigenvalues. In the QLM, the $U(1)$ gauge field is first replaced by a $SU(2)$ spin operators by identifying $E_{n,n+1}\equiv S^z_{n,n+1}$ and $U_{n,n+1}\equiv S^+_{n,n+1}$. The spin operators are then re-written as the rishon operators according to $S^+_{n,n+1}=c_{n,l}c_{n+1,r}^\dagger$ and $S^z_{n,n+1}=\frac{1}{2}(c_{n+1,r}^\dagger c_{n+1,r}-c_{n,l}^\dagger c_{n,l})$, in which $l$ and $r$ label the position of the rishon particle. In the rishon representation the Gauss law becomes
 \begin{equation}
  c_{n,r}^\dagger c_{n,r}+\psi^\dagger_n\psi_n+c_{n,l}^\dagger c_{n,l}=N-\frac{(-1)^n-1}{2}, \label{gaussrish}
\end{equation}
in which $N\equiv\mathcal{N}_{n,r}+\mathcal{N}_{n,l}$ is the total rishon number per link. It is then easy to identify gauge invariant states using occupation number basis $|\mathcal{N}_{n,r},\mathcal{N}_n,\mathcal{N}_{n,l}\rangle$ in the rishon language. (See Appendix \ref{appendix1} for the detail properties of the  local physical states.) Although the dimension of the local Hilbert space has been reduced to a finite value, QLM has been shown to demonstrate similar static and dynamic behaviors compared to the Schwinger model~\cite{Chandrasekharan,Brower,Surace,Notarnicola}.

The model has two phases: $PC$-symmetric and $PC$-broken phases. The total electric flux $\mathcal{E}\equiv\sum_n\langle S^z_{n,n+1}\rangle/2$ serves as the order parameter of the model. In this work, we first study the $S=1$ QLM, to establish the scaling behavior. We then study the $S=1/2$ QLM to examine the universality of the scaling. The phase diagram of $S=1/2$ and $S=1$ QLM has been studied using the tensor network method~\cite{Rico}. For the $S=1$ QLM, with $E_0=1/2$ and $\kappa=1/2$, the critical point is shown to be at $\mu_c=-0.2173$. On the other hand, for the $S=1/2$ QLM, with $E_0=0$ and $\kappa=1$, the critical point is located at $\mu_c=0.655$. It also has been shown that this phase transition belongs to the Ising universality class~\cite{Sachdevbook}.

\section{the KZM and the Combined topological defects in LGT} \label{combkzm}
\subsection{Brief review of the KZM}

\begin{figure*}[tbp]
	\includegraphics[angle=0,scale=0.35]{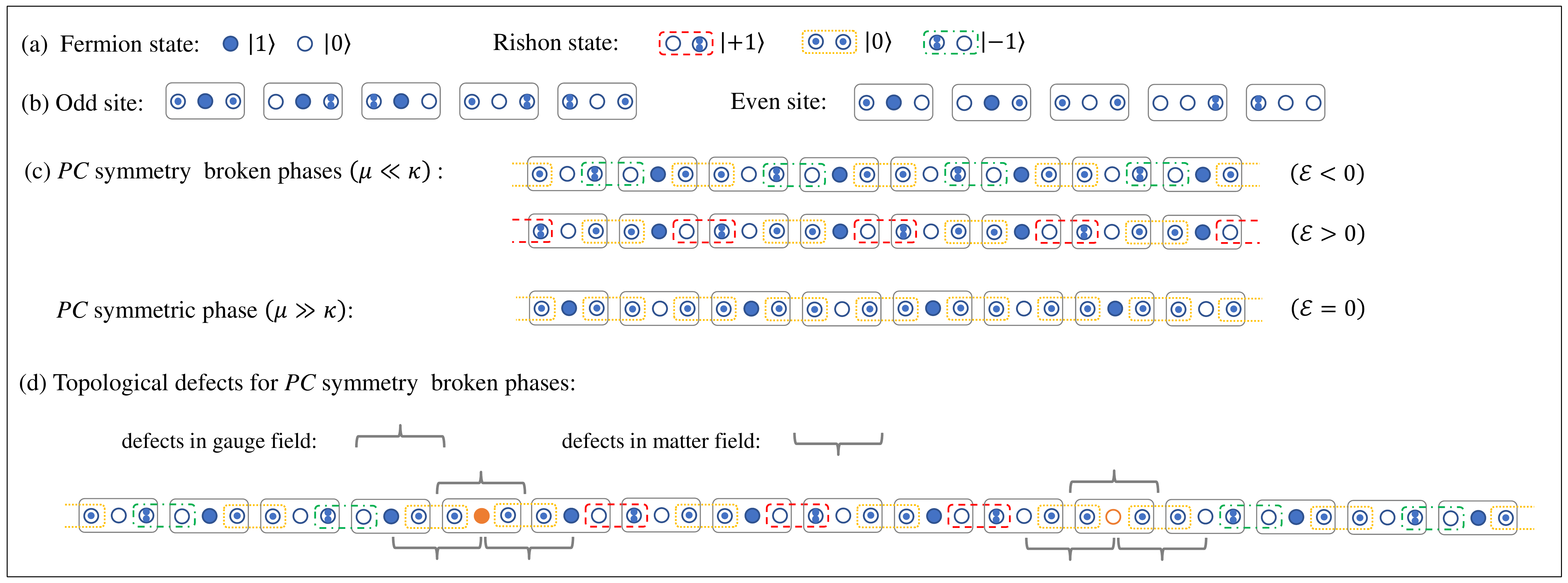}
	\caption{Ground states and topological defects of the Schwinger model in the $N=2$ quantum link representation. The pictorial representations of the fermion and the rishon states in (a) are resembled as the physical allowed local quantum states in odd and even sites (b). Two degenerate $PC$ symmetry broken states with nonzero electric flux for $\mu\ll\kappa$, and the $PC$-symmetric state with zero flux for for $\mu\gg\kappa$ are shown in (c). Two kinds of combined topological defects, made up by two fermion excitations and one gauge field excitation, in Ground state $1$ are shown in (d).}
	\label{fig1}
\end{figure*}

Topological defects are excitations emerging at the position where irreconcilably different broken-symmetry vacuum is chosen. Depending on the symmetry of the system, they can be monopoles, vortex lines, domain walls, and so on. The topological defects are energetically costly but topologically stable, since the whole manifold would have to be rearranged to get rid of the defects.

The KZM shows that the topological defects are created when a system is driven across a critical point and the density of these topological defects $n_{\rm TD}$ scales with the driving rate $R$~\cite{Kibble1,Zurek1}. Denoting the distance of the relevant parameter to the critical by $g$, here we consider the driven dynamics of the form $g(t)=g_0+Rt$. We set $g_0$ to be far from the critical point and hence is irrelevant. Typically, two time scales are involved in the KZM analysis. One is the intrinsic relaxation time scale of the system $\zeta_r$ and it scales with the energy gap $\Delta$ as $\zeta_r\sim \Delta^{-1}$. The other is the driven time scale $\zeta_d$ induced by the external driving and it scale with driving rate $R$ as $\zeta_d\sim R^{-z/r}$. Here $z$ is the dynamic exponent and $r \equiv z+1/\nu$ with $\nu$ being the standard exponent associated with the correlation length $\xi\propto |g|^{-\nu}$. Depending on the relative strength of  $\zeta_r$ and $\zeta_d$, the whole process can be separated into three stages: In the initial stage, the system is far from its critical point and $\zeta_r<\zeta_d$. Consequently, the system evolves adiabatically along its ground state.  As the system gets closer to the critical point, the system enters the impulse region in which $\zeta_r>\zeta_d$. This happens when $ g = \hat{g}_1 = c_1R^{1/\nu r}$ where $c_1$ is an irrelevant non-universal constant. The KZM assumes that in this stage the system ceases to evolve and the state remains the same as the one at $g=\hat{g}_1$ as a result of the critical slowing down. Continuing driving the system away from the critical point, eventually the system will leave the impulse region and enter into another adiabatic region at $g=\hat{g}_2=c_2R^{1/\nu r}$, where $c_2$ is another irrelevant non-universal constant. However, although $\zeta_r<\zeta_d$ in this region, the system will not evolve along its ground state. This is because the state at $\hat{g}_2$ is the one at $\hat{g}_1$, which is different from the ground state at $\hat{g}_2$. As a result, many excitations will appear. Among them, some topologically stable excitations can survive even for very long time. The KZM demonstrates that the density of the topological defects $n_{\rm TD}$ at this stage obeys the scaling
\begin{equation}
  n_{\rm TD}\propto R^{\frac{1}{r}}. \label{kzmtp}
\end{equation}
For the quantum Ising universality class, one has $r=2$ since $z=1$ and $\nu=1$. Although the freezing of the evolution in the impulse region has been shown to be an oversimplified assumption, the KZM prediction of Eq.~(\ref{kzmtp}) has been verified in various systems~\cite{Ulm,Pyka,Navon}.

\subsection{\label{combkzm2}Combined topological defects in quantum link model}

In contrast to the conventional symmetry-breaking phase transitions, the formation of the topological defects in the LGT must obey the local constraint of the Gauss' law. To explore properties of topological defects of the QLM, we first identify the physical allowed states on even site and odd site respectively. As shown in Fig.~\ref{fig1}, there are only $5$ allowed states on each site. We also sketch the representative states of the different phases at the strong coupling limit $|\mu| \gg |\kappa|$. Consider now the representative $PC$-broken state with $\mathcal{E}<0$. On odd site has one $|\mathcal{N}_{n,r}=1,\mathcal{N}_n=0,\mathcal{N}_{n,l}=2\rangle$, while on even site one has $|\mathcal{N}_{n,r}=0,\mathcal{N}_n=1,\mathcal{N}_{n,l}=1\rangle$. The topological defects will appear when one inserts a segment of another representative $PC$-broken state with $\mathcal{E}>0$. However, due to the constraint of the Gauss' law, this segment can only be assembled on the right-hand side of the even site or the left-hand side of the odd site. From Fig.~\ref{fig1}, one finds that one topological defect is always being made up of two matter field excitation and one link excitation. This is similar to the excitation in the superconductivity, in which the flux is tied up by the vortex~\cite{Hindmarsh,Stephens}.

\subsection{KZM in the quantum link model}

\begin{figure}[tbp]
\includegraphics[angle=0,scale=0.35]{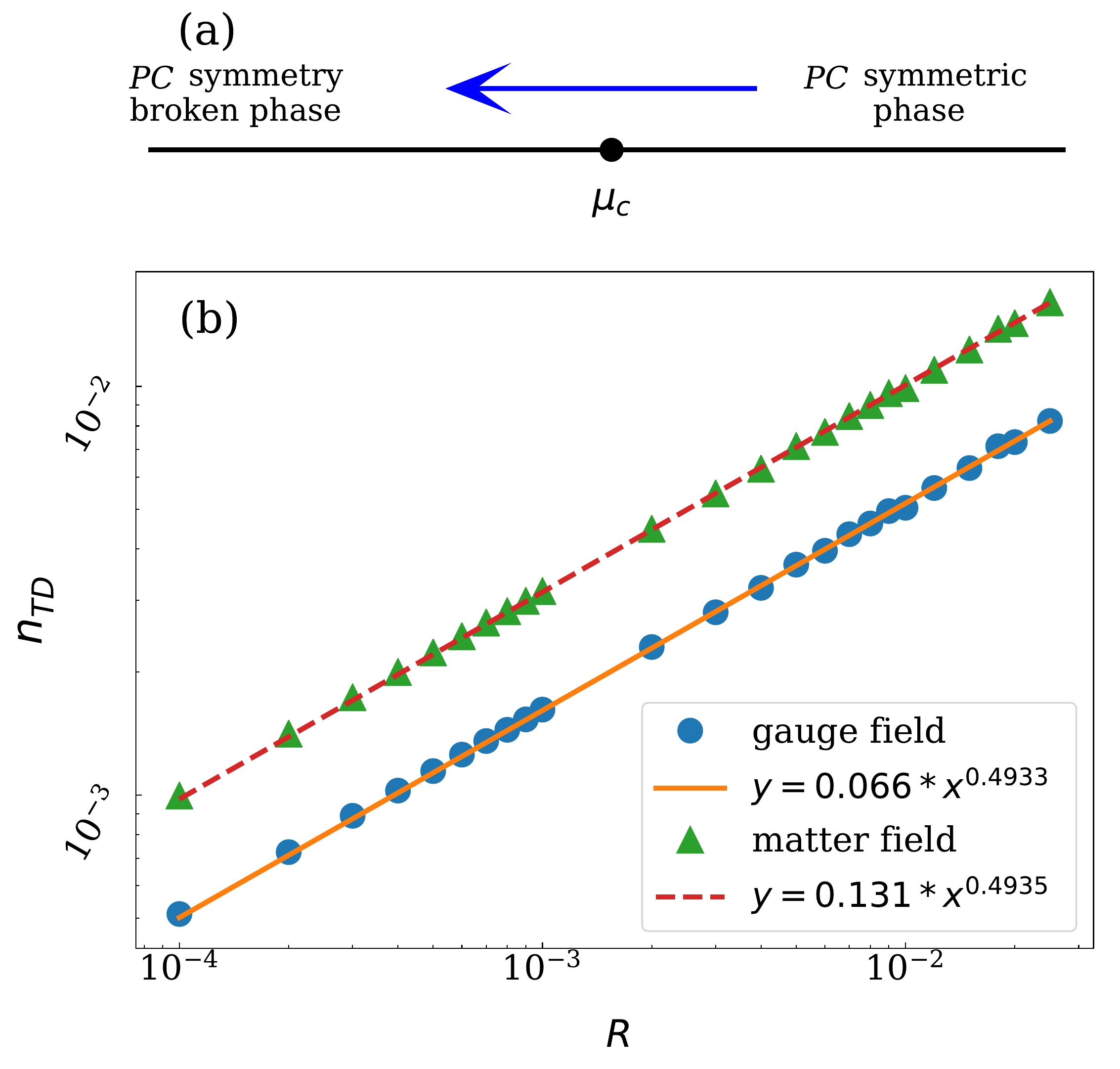}
  \caption{(a) Schematic diagram of the driven protocol of KZM for the $S=1$ QLM: $\mu(t) = \mu_0 - Rt$ until it reaches $\mu_f$.
  (b) Density of the topological defects after the quench versus the driving rate. We start from $\mu_0=0.5$ and measure the density at $\mu_f=-5$.
   Double logarithmic coordinates are used.}
\label{fig2}
\end{figure}

We first numerically explore the KZM in the QLM with $N=2$ $(S=1)$. We start from the $PC$-symmetric ground state away from the critical point, at $\mu_0=0.5 > \mu_c = - 0.2173$, and drive the chemical potential as $\mu(t) = \mu_0 - Rt$ into the $PC$-broken phase until it reaches certain final $\mu_f = -5$ that is far away from the critical point, as sketched in Fig.\ref{fig2}(a). During the whole process, the distance to the critical point is $g = \mu(t) - \mu_c$.

Similar to the domain wall in the $1$D quantum Ising model~\cite{qkz1,qkz2}, the topological defect in the gauge field is defined as
\begin{equation}
  n_{\rm TD}^G\equiv-\sum_n [\langle S_{n,n+1}^zS_{n+2,n+3}^z\rangle-\langle S^z\rangle^2_G], \label{gaugetd}
\end{equation}
where $\langle S^z\rangle_G=\sum_n\langle S^z_{n,n+1}\rangle_G/2$. Here, $\langle\rangle_G$ denotes expectation value in the ground state and $\langle\rangle$ denotes expectation value in the dynamic evolution state. Similarly the topological defect in the matter field can be defined as
\begin{equation}
  n_{\rm TD}^M\equiv\sum_n [\langle (2\mathcal{N}_n-1)(2\mathcal{N}_{n+1}-1)\rangle+\langle (2\mathcal{N}-1)\rangle^2_G],
\end{equation}
which can be obtained by inspecting Fig.~\ref{fig1}. We employ the infinite time-evolving block decimation (iTEBD) method to simulate the time evolution of the state~\cite{Vidali}. The truncation dimension is kept to be $50$ and the time difference is chosen as $0.01$. The numerical calculation was done using the Uni10 tensor network library ~\cite{uni10}. Although the translational symmetry breaks for an individual measurement, here we can detect the superposition of these defects~\cite{Dziarmagan,Gillman,Gillman1}.

Figure~\ref{fig2} shows the density of the topological defects for different rates. By power law fitting, one finds that the scaling of the density of excitations satisfied $n_{\rm TD}^{G/M}\propto R^{1/2}$ for both gauge field and matter field. Thus the KZM of Eq.~(\ref{kzmtp}) in the QLM is verified. Moreover, we find that the ratio of the coefficients for the gauge field and the matter field is about $1/2$. This confirms our discussion in Sec.~\ref{combkzm2}.

\section{The FTS in the quantum link model} \label{ftss}
In the impulse region, the KZM states that the system does not evolve. However, it has been shown that this is an oversimplified assumption. The FTS improves the understanding of the critical dynamics in the impulse region by demonstrating that the system evolves in the impulse region according to the characteristic time scale $\zeta_d\sim R^{-z/r}$. Accordingly, the evolution of the macroscopic quantities should satisfy the full scaling forms. For example, the electric flux $\mathcal{E}$ should obey~\cite{Zhong1,Zhong2,qkz10,Yin}
\begin{equation}
    \mathcal{E}(g,R) = R^{\beta/\nu r} f_1 (g R^{-1/ \nu r}),
    \label{fluxfts}
\end{equation}
similar to the FTS scaling of the order parameter. In Eq.~(\ref{fluxfts}), $\beta=1/8$ for the $1$D quantum Ising universality class~\cite{Sachdevbook}.

\begin{figure}[tbp]
\includegraphics[angle=0,scale=0.27]{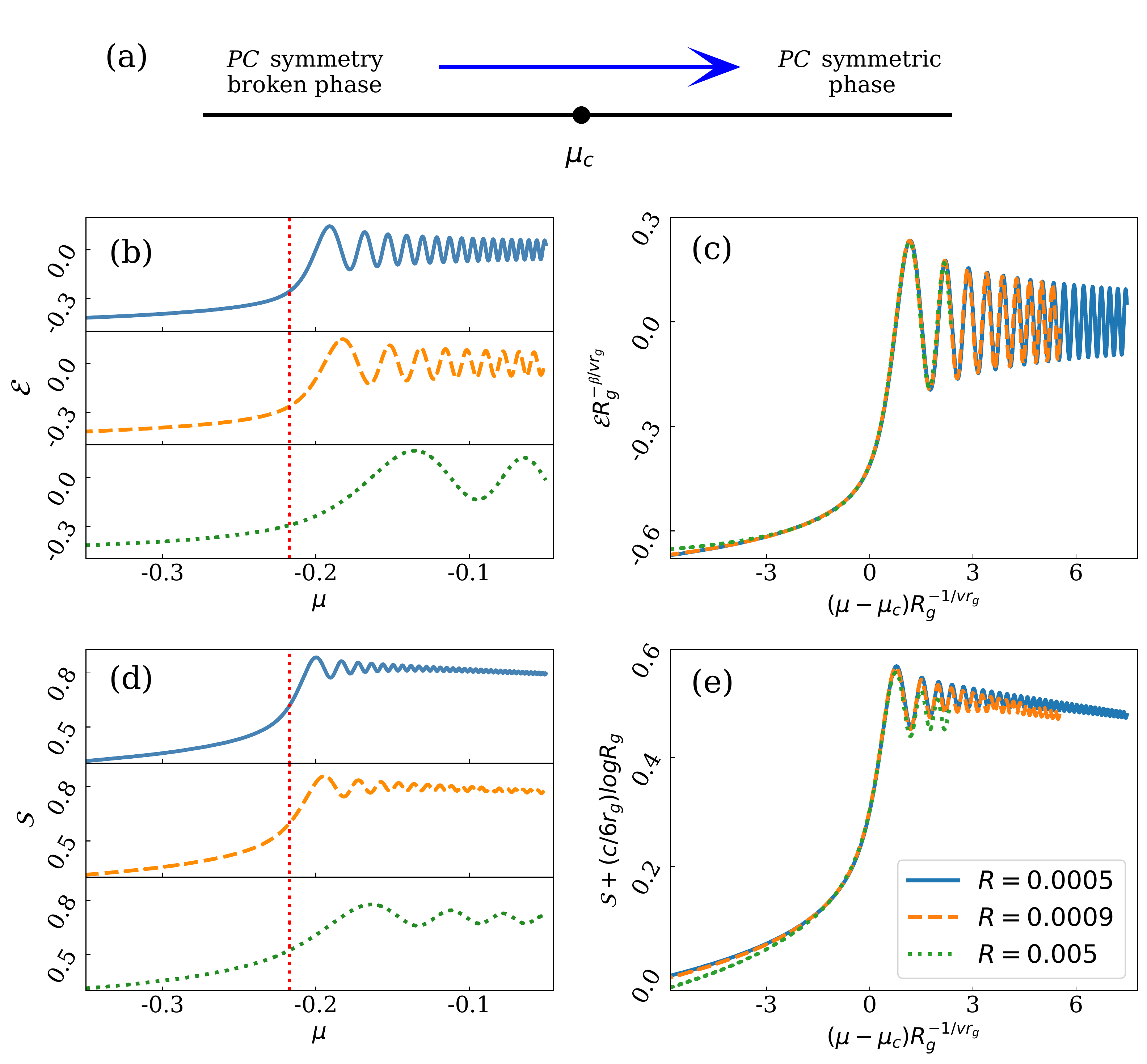}
  \caption{
  (a) Schematic diagram of the driven protocol of FTS for the $S=1$ QLM: $\mu(t) = \mu_0 + Rt$.
  (b) Electric flux $\mathcal{E}$ versus $\mu$ with three different rates. (c) Rescaled curves of $\mathcal{E}$.
  (d) Entanglement entropy $S$ versus $\mu$ with three different rates. (e) Rescaled curves of $S$.
   All the curves are started at $\mu_0=-3$. The vertical dotted lines denote the position of the critical point.}
\label{fig3}
\end{figure}

Besides the electric flux, the evolution of the von-Neumann entanglement entropy also demonstrates a scaling behavior. The von-Neumann entanglement entropy is measured as $\mathcal{S}=-\textrm{Tr}(\rho\textrm{log}\rho)$, where $\rho$ is the reduced density matrix of half of the system. For a $1$D system near its quantum critical point it has been shown that the entanglement entropy scales as $\mathcal{S}=(c/6)\textrm{log}\xi$~\cite{Eisert,Amico,Osterloh,Laorencie,Calabrese}, where $c$ is the central charge and $\xi$ is the correlation length. For the Ising universality class, $c=1/2$. In passing we note that recently, the entanglement entropy has been measured in experiments~\cite{Islam}. Since the scaling form of the correlation length $\xi$ under external driving is $\xi(g,R)=R^{-1/r}f_2(gR^{-1/\nu r})$, the entanglement entropy $S$ satisfies
\begin{eqnarray}
\begin{aligned}
\mathcal{S}(g,R)=-\frac{c}{6r}\textrm{log}R+f_3(gR^{-1/\nu r}), \label{eesc}
\end{aligned}
\end{eqnarray}
in which $f_3=-(c/6r)\textrm{log}f_2$.

To verify the FTS, we start from the $PC$-broken state at  $\mu_0=-3$ and drive the chemical potential as $\mu(t)=\mu_0+Rt$ into the $PC$-symmetric phase, as sketched in Fig.~\ref{fig3}(a). We focus on the scaling of electric flux and entanglement entropy when $\mu$ is near critical point $\mu_c$. In Fig.~\ref{fig3}(b) and (d) we show $\mathcal{E}$ and $S$ versus $\mu$ respectively with three different rate $R$. We have checked that the results are independent of the particular $\mu_0$ used, as long as $\mu_0$ is sufficient far away from the critical point. In Fig.~\ref{fig3}(c) and (e) we verify that curves with different $R$ will collapse into a single one if the scaling form Eq.~\ref{fluxfts} or Eq.~\ref{eesc} is used. These results demonstrate that the FTS is applicable for the phase transition in the QLM.

\section{Universality of KZM and FTS in the quantum link model} \label{universality}

\begin{figure}[tbp]
\includegraphics[angle=0,scale=0.35]{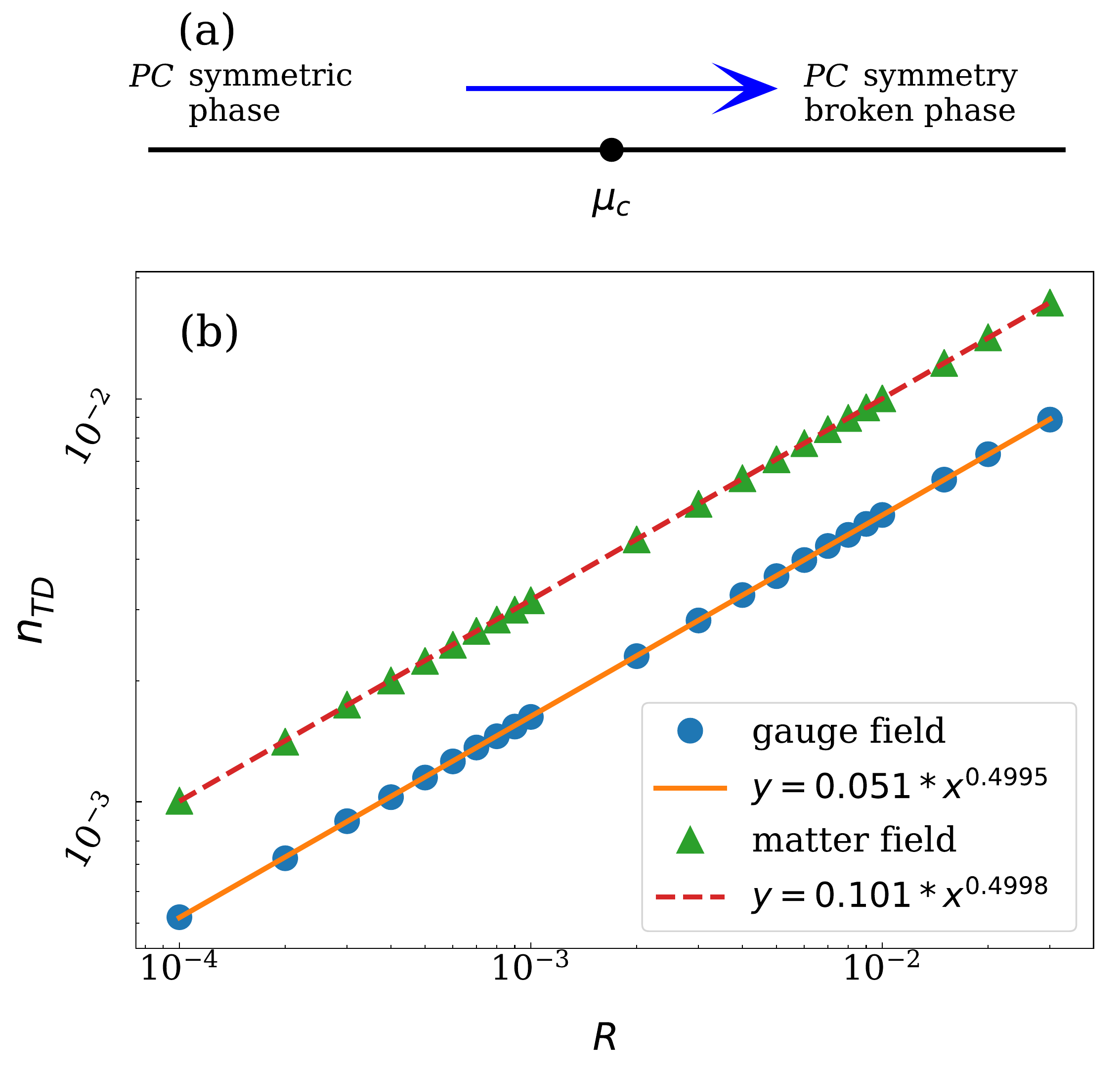}
\caption{(a) Schematic diagram of the driven protocol of KZM for the $S=1/2$ QLM: $\mu(t) = \mu_0 + Rt$ until it reaches $\mu_f$.
  (b) Density of the topological defects after the quench versus the driving rate. We start from $\mu_0=0$ and measure the density at $\mu_f=5$.
   Double logarithmic coordinates are used.}
\label{fig6}
\end{figure}

\begin{figure}[tbp]
\includegraphics[angle=0,scale=0.27]{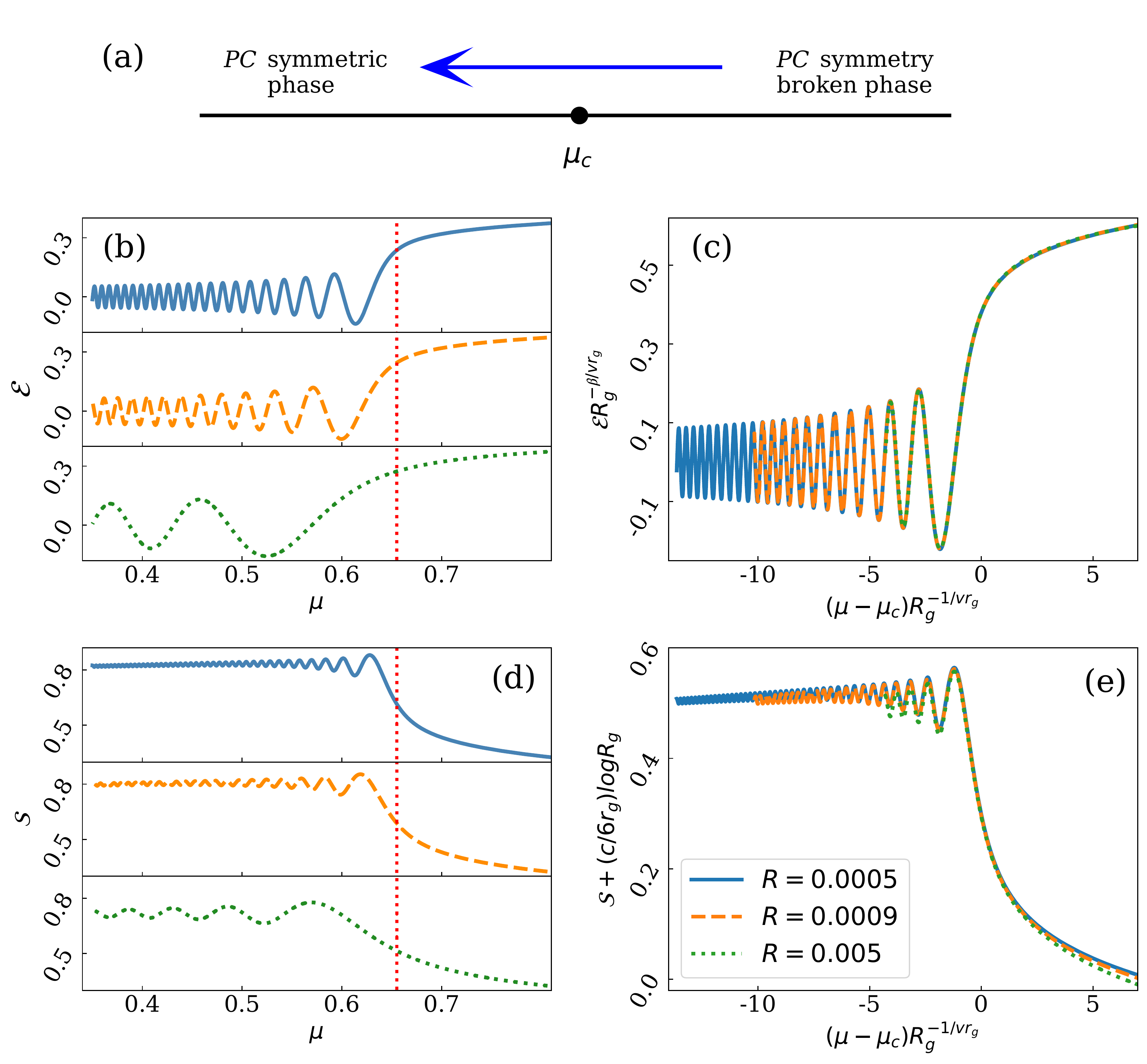}
\caption{
  (a) Schematic diagram of the driven protocol of FTS for the $S=1/2$ QLM: $\mu(t) = \mu_0 - Rt$.
  (b) Electric flux $\mathcal{E}$ versus $\mu$ with three different rates. (c) Rescaled curves of $\mathcal{E}$.
  (d) Entanglement entropy $S$ versus $\mu$ with three different rates. (e) Rescaled curves of $S$.
   All the curves are started at $\mu_0=3$. The vertical dotted lines denote the position of the critical point.}
\label{fig7}
\end{figure}

In this section, we show that our results discussed above are independent of the choice of the rishon representation. To be concrete, we choose to verify the universality using $S=1/2$ QLM. In this case, $\mu > \mu_c$ corresponds to the symmetric phase while $\mu < \mu_c$ corresponds to the symmetry-broken phase. Furthermore, the definitions for the matter field defect and the gauge field defect are modified as
\begin{equation}
  n_{\rm TD}^M\equiv \sum_{n}[\langle(\mathcal{N}_n-\frac{1}{2})(\mathcal{N}_{n+1}-\frac{1}{2})\rangle+\langle\mathcal{N} -\frac{1}{2}\rangle^2_G], \label{mattertd1}
\end{equation}
and
\begin{equation}
  n_{\rm TD}^G \equiv -\sum_{n} ( \langle S^z_{n,n+1} S^z_{n+1,n+2} \rangle - \langle S^z \rangle^2_G ), \label{gaugetd1}
\end{equation}
respectively.

We first test the universality of the KZM in QLM. Similar analyses show that for the case of $S=1/2$ QLM, there are also two kinds of topological defects between two symmetry breaking ground states (See Fig.~\ref{fig:sketchS05} in Appendix \ref{appendix1}). Also, each topological defect is made up by one gauge field excitation and two matter field excitations. Figure~\ref{fig6} confirms the KZM of Eq.~(\ref{kzmtp}) for both the matter field defects~(\ref{mattertd1}) and the gauge field defects~(\ref{gaugetd1}). Moreover, one finds that the ratio between the coefficients of the topological defects for the gauge field and the matter field is $1/2$. This is same as the results abstained for the $S=1$ QLM as shown in Fig.~\ref{fig2}.

We then examine the FTS for $S=1/2$ QLM. Under the driven protocol as displayed in Fig~\ref{fig7}(a), the evolution of the electric flux and the entanglement entropy are shown in Fig.~\ref{fig7}(b) and (d), respectively. In Fig.~\ref{fig7}(c), one finds that the rescaled curves of $\mathcal{E}$ versus $\mu$ collapse into a single curve, confirming Eq.~(\ref{fluxfts}). In addition, Figure~\ref{fig7} (e) verifies Eq.~(\ref{eesc}) by showing the merging of the rescaled curves of $\mathcal{S}$ versus $\mu$ for various $R$ near the critical point. These results confirm the universal property of the FTS in the QLM.

\section{\label{sum}Summary and discussion}
We study the driven critical dynamics in the QLM, which describes a $U(1)$ LGT. We focus on the effects induced by the local gauge constraints of the Gauss' law. We find that the topological defects generated in the driven process are combined topological excitations. These combined topological defects are shown to be made up by both the gauge field and matter field excitations. We also shown that the density of these combined topological defects satisfies the usual KZM. Moreover, we study the scaling behavior in the whole driven process and demonstrates that both the electric flux and the entanglement entropy satisfy the FTS theory.

Some remarks are added here. (a) This model has a sister model in which the external field is uniform~\cite{Coleman,Byrnes}. An Ising phase transition also happens therein for changing $\mu$. It is expected our present results should be applicable therein. (b) In Ref.~\onlinecite{Hindmarsh,Stephens}, the gauge field is gapless and its initial state can be remembered for a very long time scale. Thus the topological defects are affected by these modes. However, for our present case, the gauge field fluctuation is gapped in the initial stage. So, the topological defects are only affected by the Ising phase transition. (c) Recently, quantum simulators for both Abelian and non-Abelian LGTs have been proposed based on the cold atom systems. Not only the ground state properties but also the real time dynamics can be manipulated and detected in these systems~\cite{Martinez,Banerjee,Hebenstreit,Spitz}. In particular, the string dynamics in the $S=1/2$ QLM has been realized in a recent experiment~\cite{Surace}, in which the Rydberg blockade mechanism is employed to simulate the local constraint of the Gauss' law and the slow relaxation dynamics is observed. Moreover, the KZM and FTS are also verified in the Rydberg atomic experiment~\cite{Keesling}, in which tunable Rydberg blockade can be recast to a quantum clock model. Thus, it is expected that our results could be examined in similar experiments.

\section{\label{Ack}Acknowledgments}
We acknowledge the support by Ministry of Science and Technology (MOST) of Taiwan through Grant No. 107-2112-M-007-018-MY3. We also acknowledge the support from the National Center for Theoretical Science (NCTS) of Taiwan. SY is supported in part by China Postdoctoral Science Foundation (Grant No. 2017M620035).

\appendix

\section{Gauge invariant states and effective Hamiltonian}\label{appendix1}
In this work, we combine the occupation numbers of the rishon particles on the right $(x,r)$, the fermion $(x)$, and the rishon particles on the left $(x,l)$ as the local computational states $|\mathcal{N}_{n,r},\mathcal{N}_n,\mathcal{N}_{n,l}\rangle$~\cite{Rico}.

\subsection{$S=1$ quantum link model}

For $S=1$ QLM, each rishon state has two particles.
To constraint the rishon  number, an additional interacting term with large positive $U$ is added to the Hamiltonian (\ref{modelham}).
In this case, we choose $E_0=1/2$, the effective Hamiltonian should be rewritten as
\begin{eqnarray}
\begin{aligned}
H_{\rm spin-1}=& \frac{1}{4}\sum_n\left[c_{n+1,r}^\dagger c_{n+1,r}-c_{n,l}^\dagger c_{n,l} - (-1)^n\right]^2  \\
& - \kappa\sum_n\left(\psi_n^\dagger c_{n,l}c_{n+1,r}^\dagger\psi_{n+1}+{\rm H.c.}\right)  \\
& + \mu\sum_n(-1)^n\psi_n^\dagger\psi_n \\
& + U\sum_n\left(c_{n+1,r}^\dagger c_{n+1,r} + c_{n,l}^\dagger c_{n,l} - 2\right)^2 .
\label{eq:hamS10}
\end{aligned}
\end{eqnarray}

On the other hand, due to the constraint condition (\ref{gaussrish}), only five the gauge invariant states are allowed in both odd site and even site, as shown in Fig.~\ref{fig1}(b).

When $\mu\gg \kappa$, due to the gauge invariance, only one ground state is possible for the Hamiltonian (\ref{eq:hamS10}). The electric flux is zero in this ground state, which is invariant under $P$ and $C$ transformation.

When $\mu\ll\kappa$, $P$ and $C$ symmetry breaking occurs and $\mathcal{E}\neq0$, resulting in two degenerate ground states.
The topological defects happen if one inserts a segment of ground state with $\mathcal{E}>0$ into the ground state with $\mathcal{E}<0$, similar to the domain wall in the 1D quantum Ising model.
The topological defects in gauge field and matter field are sketched in Fig.~\ref{fig1}(d).

\subsection{$S=1/2$ quantum link model}

\begin{figure*}[tbp]
	\includegraphics[angle=0,scale=0.35]{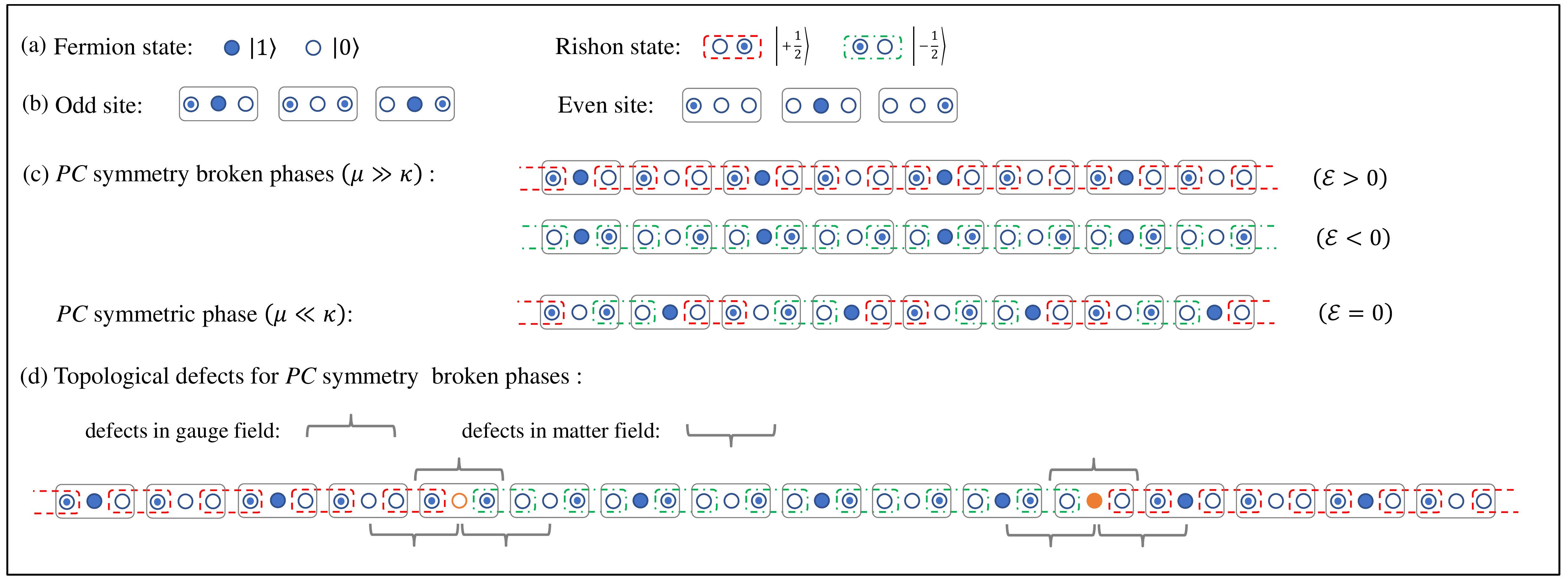}
	\caption{Ground states and topological defects of the Schwinger model in the $N=1$ quantum link representation. The pictorial representations of the fermion and the rishon states in (a) are resembled as the physical allowed local quantum states in odd and even sites (b). (c) Two degenerate $PC$ symmetry broken ground states with nonzero electric flux, and one $PC$-symmetric ground state with zero flux. Two kinds of combined topological defects, made up by two fermion excitations and one gauge field excitation are shown in (d).}
	\label{fig:sketchS05}
\end{figure*}

For $S=1/2$ QLM, each rishon state has one particle.
Similarly, an additional interacting term with large positive $U$ is added to the Hamiltonian (\ref{modelham}), and the effective Hamiltonian is given by follow:
\begin{eqnarray}
\begin{aligned}
H_{\rm spin-\frac{1}{2}} = & \frac{1}{4}\sum_n\left[c_{n+1,r}^\dagger c_{n+1,r}-c_{n,l}^\dagger c_{n,l} \right]^2  \\
& - \kappa\sum_n\left(\psi_n^\dagger c_{n,l}c_{n+1,r}^\dagger\psi_{n+1}+{\rm H.c.}\right)  \\
& + \mu\sum_n(-1)^n\psi_n^\dagger\psi_n \\
& + 2U\sum_n\left(c_{n+1,r}^\dagger c_{n+1,r} - \frac{1}{2}\right)\left( c_{n,l}^\dagger c_{n,l} - \frac{1}{2}\right).
\label{eq:hamS05}
\end{aligned}
\end{eqnarray}

In this case, due to the constraint condition (\ref{gaussrish}), three the gauge invariant states are allowed in both odd site and even site, as shown in Fig.~\ref{fig:sketchS05}(b).

When $\mu\ll\kappa$, the electric flux $\mathcal{E}=0$ and the ground state is invariant under $P$ and $C$ transformation; when $\mu\gg\kappa$, $P$ and $C$ symmetry breaking occurs and $\mathcal{E}\neq0$.
The topological defects in gauge field and matter field are sketched in Fig.~\ref{fig:sketchS05}(d).

\section{Infinite matrix product state and quantum number}

\begin{figure}[tbp]
	\includegraphics[angle=0,scale=0.34]{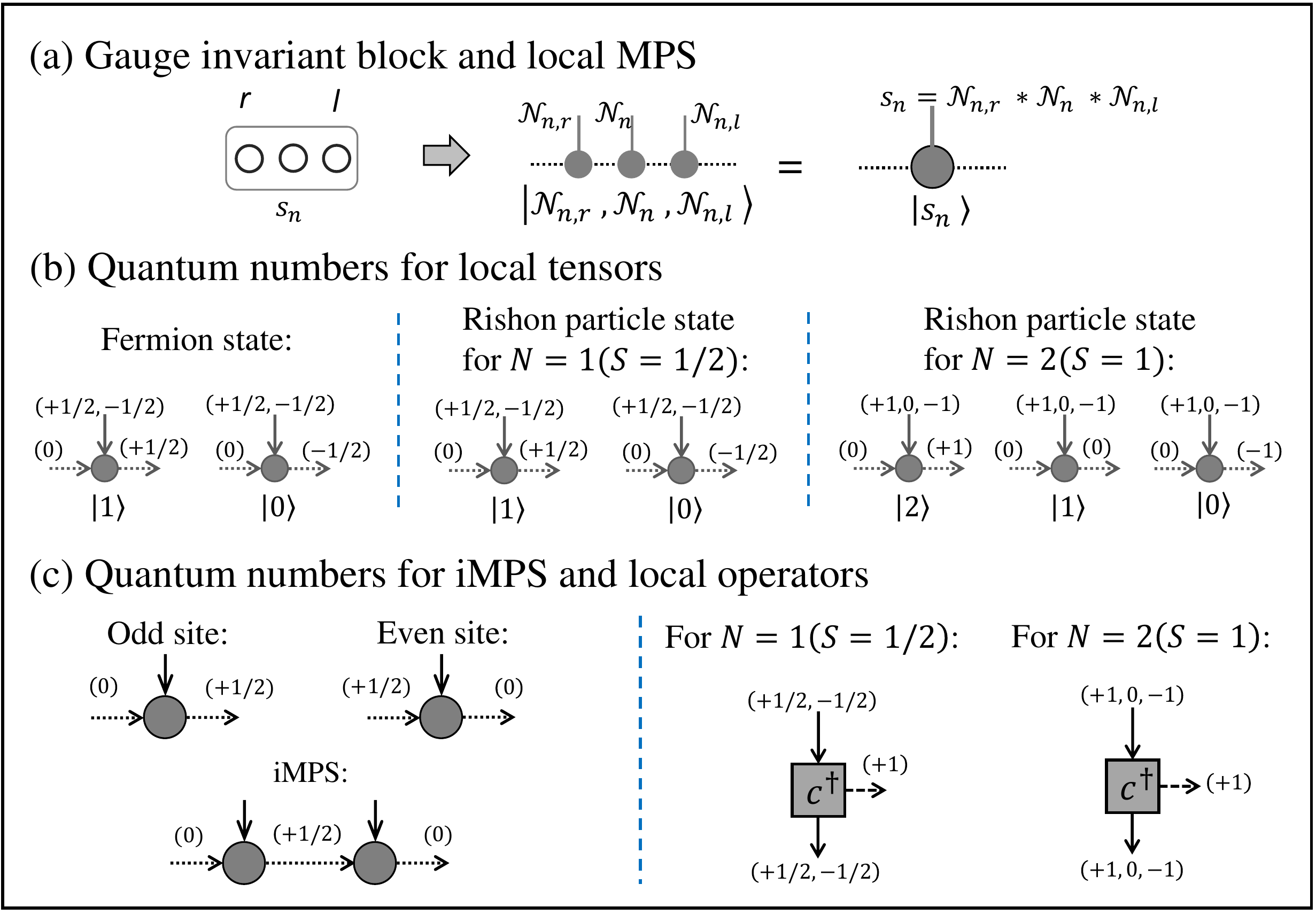}
	\caption{(a) Local gauge invariant block as the computational states. (b) The quantum numbers of $U(1)$ symmetry for local fermion states and rishon particle states. The quantum numbers of a bond are noted in the parenthesis. (c) The diagrammatic notations of the iMPS and the rishon operators. The constraint (\ref{gaussrish}) is fulfilled by fixing the quantum number in the virtual bond of iMPS.  An auxiliary bond with quantum number $+1$, represented as dashed lines, is applied to the rishon operator $c^\dagger$.
	\label{fig:mps}}
\end{figure}

In this work, the local physical states $|\mathcal{N}_{n,r},\mathcal{N}_n,\mathcal{N}_{n,l}\rangle$ are considered as the local computational states, which is sketched in Fig.~\ref{fig:mps}(a).
Thus, the ground state can be expressed in the form of an infinite matrix product state (iMPS),
\begin{eqnarray}
\begin{aligned}
|\Psi\rangle_G &= \sum_{\bf s} \cdots\lambda^{s_{odd}}\Gamma^{s_{odd}}\lambda^{s_{even}}\Gamma^{s_{even}}\cdots|\bf s\rangle \\
	&=\sum_{\bf s} \cdots A^{s_{odd}} A^{s_{even}}\cdots|\bf s\rangle.
\end{aligned}
\end{eqnarray}
To fulfill the constraint (\ref{gaussrish}), this iMPS has to be restricted.

We realize the constraint (\ref{gaussrish}) by fixing the quantum number of $U(1)$ symmetry.
Figure \ref{fig:mps}(b) illustrates the graphical notations of the $U(1)$-symmetric matrix product states for the local fermion states and rishon particle states.
And the tensor network diagrams for rishon operators are graphically  shown in Fig.~\ref{fig:mps}(c).
In this representation, for $S=1$ QLM, the quantum number would add $(+1/2)$ at the odd-site MPS, and $(-1/2)$ in the even-site MPS.
As a result, the gauge constraint (\ref{gaussrish}) is fulfilled if one sets the quantum numbers in the virtual bond of iMPS as shown in Fig.~\ref{fig:mps}(c).

For $S=1/2$ QLM, the same conclusion can be drawn from a similar analysis.


\end{document}